\RequirePackage{fix-cm}
\documentclass[smallextended]{svjour3}       % onecolumn (second format)
\smartqed  % flush right qed marks, e.g. at end of proof
\usepackage{graphicx}
\begin{document}

\title{Strain effects on the ionic conductivity of Y-doped ceria: A simulation study}



\author{Mario Burbano  \and Dario Marrocchelli \and Graeme W. Watson}


\institute{M. Burbano \at
              School of Chemistry and CRANN \\
              Trinity College Dublin \\
              Dublin 2 \\
              Ireland \\
              Tel.: +353-877698113 \\
              \email{burbanom@tcd.ie}           %  \\
           \and
           D. Marrocchelli \at
              School of Chemistry and CRANN \\
              Trinity College Dublin \\
              Dublin 2 \\
              Ireland \\
              Department of Materials Science and Engineering \\
	      Department of Nuclear Science and Engineering \\
              Massachusetts Institute of Technology \\
		Cambridge, MA \\
		USA \\
              \email{dmarrocc@mit.edu}
						\and
           G.W. Watson \at
           		School of Chemistry and CRANN \\
           		Trinity College Dublin \\
           		Dublin 2 \\
              Ireland \\
              Tel.: +353-18961357 \\
              \email{watsong@tcd.ie}
}

\date{Received: date / Accepted: date}

\maketitle

\begin{abstract}
In this paper we report a computational study of the effects of strain on the conductivity of Y-doped ceria (YDC).
%This computational study seeks to address conflicting literature reports regarding the conductivity changes undergone by fluorite-structured oxide ion conductors under isotropic and anisotropic tensile strain \cite{Garcia-BarriocanalEtAl_Science2008,Guo_Science2009,Kilner_NM2008,KushimaAndYildiz_JMC2010,PergolesiEtAl_ACSN2012,SannaEtAl_AFM2009}. 
This material was chosen as it is of technological interest in the field of Solid Oxide Fuel Cells (SOFCs). The simulations were performed under realistic operational temperatures and strain ($\epsilon$) levels. For bulk and thin film YDC, the results show that tensile strain leads to conductivity enhancements of up to 3.5\,$\times$ and 1.44\,$\times$, respectively. The magnitude of these enhancements is in agreement with recent experimental and computational evidence. In addition, the methods presented herein allowed us to identify enhanced ionic conductivity in the surface regions of YDC slabs and its anisotropic character.

\end{abstract}

\section{Introduction}
\label{intro}
Solid Oxide Fuel Cells (SOFCs) are electrochemical devices which efficiently generate electricity from a variety of fuels \cite{Ormerod_CSR2003}. As such, they have the potential to play a key role in the energy production landscape in years to come \cite{WachsmanEtAl_EES2012,WachsmanAndLee_S2011}. Significant research efforts have been focused, over the past decades, on improving the performance of the oxide ion conductors used either as electrolytes or cathodes in these devices \cite{Ormerod_CSR2003,Steele_SSI2000a,Goodenough_ARMR2003}. Operating SOFCs at temperatures lower than those required by current Yttria Stabilized Zirconia (YSZ) technologies ($\sim$\,1273\,K) would allow for the use of cheaper materials in the fabrication of the cells, as well as, facilitating their possible use in portable electronics, e.g. $\mu$-SOFCs \cite{Rupp_SSI2012}. In particular, Rare Earth doped ceria (REDC), Ce$_{1-x}$RE$_{x}$O$_{2-x/2}$, where RE = Y (YDC), La (LDC), Nd (NDC), Sm (SDC), Gd (GDC), etc has been studied extensively given its relatively high ionic conductivity ($\sigma_{i}$) in the intermediate temperature range (873\,K -- 1073\,K). This process is known to occur by means of vacancy migration in the anion sublattice \cite{TullerAndNowick_JES1975,TullerAndNowick_JPCS1977}. Equation \ref{eq:vac_form} illustrates, in Kr\"oger-Vink notation, the formation of charge compensating vacancies ($V^{\cdot\cdot}_{O}$) upon substitution of a host Ce$^{4+}$ cation by a dopant RE$^{3+}$ cation ($\textrm{RE}^{'}_{\textrm{Ce}}$). 

\begin{equation}
\mathrm{RE_{2}O_{3} + (CeO_{2})_{n} \rightarrow (CeO_{2})_{n-2} + 2RE^{'}_{Ce} + 3O^{x}_{O}} + V^{\cdot\cdot}_{O}
\label{eq:vac_form}
\end{equation} 

The interactions between these defects can be classified as cation-vacancy, vacancy-vacancy and cation-cation \cite{BogicevicEtAl_PRB2001,BogicevicAndWolverton_PRB2003,NavrotskyEtAl_FD2007,PietrucciEtAl_PRB2008,Navrotsky_JMC2010,NorbergEtAl_CM2011,MarrocchelliEtAl_CM2011,BurbanoEtAl_CM2012}.  In practice, these interactions can prove deleterious to the ionic conductivity because they decrease the number of mobile vacancies available, hence the need for high operating temperatures \cite{ChenEtAl_CM2012}. \newline

From its outset, much of the research within the context of electrolyte applications has focused on the optimization of ionic conductivity from a compositional point of view, i.e. the identification of the dopant species and the concentration at which defect interactions are minimized, especially those between dopants and vacancies \cite{BogicevicAndWolverton_PRB2003,ButlerEtAl_SSI1983,Kilner_SSI1983,BalazsAndGlass_SSI1995,HayashiEtAL_SSI2000,AnderssonEtAl_PNAS2006}. In the case of ceria, these studies have identified RE elements, such as Gd$^{3+}$, Y$^{3+}$, Sm$^{3+}$ and Pm$^{3+}$ as the best candidate dopants, given that their radius mismatch with the host cation (Ce$^{4+}$) balances the competing electrostatic and elastic components of the defect interactions which control their association \cite{WangEtAl_AC2011}. Nevertheless, recent studies have shown that in the limit where cation-vacancy interactions are reduced to a minimum, it is vacancy-vacancy association which ultimately determines the ionic conductivity drop as a function of dopant concentration in fluorite-structured materials, such as, YDC \cite{BurbanoEtAl_CM2012}, YSZ and Scandia Stabilized Zirconia (ScSZ) \cite{MarrocchelliEtAl_CM2011}.  This means that further improvements in the performance of IT-SOFC electrolytes must come from the exploitation of different optimization routes \cite{WachsmanEtAl_EES2012,WachsmanAndLee_S2011,Rupp_SSI2012}.  \newline

Current thin film deposition techniques allow for the fabrication of epitaxially grown oxides with high levels of control over the microstructure, stoichiometry and lattice mismatch of individual layers \cite{FabbriEtAL_STAM2010,SantisoAndBurriel_JSSE2011,OrsiniEtAl_SM2009,SannaEtAl_AFM2009,SannaEtAl_Small2010}. These developments have attracted much attention to this area of research because they might open a new avenue for SOFC electrolyte optimization in a way similar to that which has been achieved in Si-based semiconductors \cite{ChuEtAl_ARMR2009}. Changes in the ionic conductivity of thin film electrolytes, whether detrimental or beneficial, are ascribed to interfacial effects present at phase boundaries. One such effect is the formation of space charge regions which have been shown to cause a dramatic increase in the fluoride ion conductivity of BaF$_{2}$/CaF$_{2}$ heterostructures \cite{SataEtAl_N2000,SayleEtAl_CC2003}. However, the the Debye screening length of extrinsic ionic conductors, such as, doped ceria/zirconia, is too short (given that it is inversely proportional to the square root of the defect concentration) and, as a result, space charge regions are not expected to play an important role in these systems \cite{FabbriEtAL_STAM2010,AzadEtAl_APL2005,SannaEtAl_Small2010}. Coherent growth of a thin film on a substrate, or heterostructure, leads to \textit{strain} ($\epsilon$) due to the structural mismatch which arises from differences in their lattice vectors (Equation \ref{eq:strain}). 

\begin{equation}
\epsilon = \frac{a_{1}- a_{0}}{a_{0}}
\label{eq:strain}
\end{equation}

\noindent Thin films can elastically accommodate strains of $\epsilon \sim$ 0.03, depending on the thickness of the film and the elastic properties of the material \cite{SantisoAndBurriel_JSSE2011,KorteEtAl_PCCP2008}. Strain levels beyond this value are typically released by the films through the formation of dislocation networks, defect clustering at the interface and rotations with respect to the substrate or other layers \cite{ChenEtAl_APL2003,SayleEtAl_JACS2002,SayleEtAl_MS2002}. \newline

Substantial research efforts have been devoted, over the last decade, to the elucidation of how interfacial effects alter ionic conductivity. Many such studies have focused on YSZ as it is the prevalent electrolyte in SOFCs. The results reported in the literature for this material show significant scatter in the range of conductivities achieved through the formation of interfaces. Experimental investigations have observed conductivity enhancements varying from a colossal 10$^{8}$ increase in $\sigma_{i}$ \cite{Garcia-BarriocanalEtAl_Science2008} down to neglegible changes \cite{PergolesiEtAl_ACSN2012} and myriad values in between \cite{KorteEtAl_PCCP2008,KosackiEtAl_SSI2005,AzadEtAl_APL2005,SannaEtAl_Small2010,SchichtelEtAl_PCCP2009,SillassenEtAl_AFM2010,SayleEtAl_JMC2006,Guo_Science2009,Guo_SM2011}. Early theoretical work using Molecular Dynamics (MD) simulations found only a small enhancement in $\sigma_{i}$ for YSZ mediated by a lowering of the activation barrier \cite{SuzukiEtAl_APL1998}, while more recent Density Functional Theory (DFT) calculations have predicted a maximum enhancement of up to four orders of magnitude in this property, at 400~K for relatively high strains ($\epsilon =$ 4\%) \cite{KushimaAndYildizECST2009,KushimaAndYildiz_JMC2010}. \newline

In the case of REDC, the conductivity enhancements observed have been more modest. For example, Chen \textit{et al.} studied thin GDC films grown on MgO (lattice mismatch of 28\%) and found only a small increase in the ionic conductivity with respect to the bulk system  \cite{ChenEtAl_APL2003}. Suzuki \textit{et al.} \cite{SuzukiEtAl_SSI2002} spin coated sapphire substrates with CeO${_2}$ thin films, as well as, GDC films of different concentrations. Their study found that the conductivity increased with diminishing film thickness. However, this change may be attributed to a lowering of the oxygen vacancy formation energy which leads to reduction of Ce$^{4+}$ to Ce$^{3+}$ (Equation \ref{eq:reduction}), and thus, to electronic conductivity \cite{Tuller_SSI2000}. 

\begin{equation}
\mathrm{2Ce_{Ce}^{x} + O_{O}^{x} \leftrightarrow V^{\cdot\cdot}_{O} + \frac{1}{2}O_{2} + 2Ce^{'}_{Ce} }
\label{eq:reduction}
\end{equation} 

\noindent This is an undesirable effect in SOFC electrolytes as it causes an internal short circuit and cell delamination through lattice expansion \cite{AtkinsonAndRamos_SSI2000,SantisoAndBurriel_JSSE2011,HullEtAl_JSSC2009,SteeleAndHeinzel_N2001,MarrocchelliEtAl_AFM2012,MarrocchelliEtAl_PCCP2012,BishopEtAl_EES2013}. Similar results were obtained by Perkins \textit{et al.}, who saw the formation of discrete micro-domains containing Ce$^{3+}$ in SDC/CeO$_{2}$ heterostructures ($\epsilon \sim$ 0.0035) grown on MgO substrates \cite{PerkinsEtAL_AFM2010}. Studies of the ionic conductivity in epitaxial SDC films grown on MgO substrates using SrTiO$_{3}$ buffer layers ($\epsilon \sim$ 0.016) found a conductivity of 0.07 S\,cm$^{-1}$ at 973\,K compared to $\sim$0.02 S\,cm$^{-1}$ for dense polycrystalline pellets at the same temperature \cite{SannaEtAl_AFM2009}. In the case of STO-buffered SDC/YSZ heterostructures ($\epsilon \sim$ 0.055)) grown on MgO, the same group found a conductivity increase of two orders of magnitude with respect to SDC polycrystaline pellets and about one order of magnitude increase compared to either SDC or YSZ thin films \cite{SannaEtAl_Small2010}. A recent computational study, which employed static calculations with interatomic potentials, predicted an enhancement in $\sigma_{i}$ of up to four orders of magnitude when tensile strain ($\epsilon$ = 0.040) was applied to bulk CeO$_{2}$ \cite{DeSouzaEtAl_EES2012}. \newline

This brief survey of our current understanding of interfacial effects highlights the need for further investigation in this area of materials research. To this end, computer simulations can play an important role because the various processes which underlie complex phenomena can be treated in a direct and controlled manner, thus, making it possible to evaluate their contribution to the overall changes observed in experiments. In this paper we have assessed how the ionic conductivity in YDC is modified when bulk and thin films are subjected to isotropic and anisotropic biaxial strain, respectively. Our study has a series of distinctive features that set it apart from previous work. Firstly, we use molecular dynamics simulations coupled with accurate dipole-polarizable interatomic potentials derived directly from \textit{ab initio} calculations \cite{BurbanoEtAl_JPCM2011,BurbanoEtAl_CM2012}.  Secondly, our use of the slab method allows us to account for the relaxation perpendicular to the plane where strain is applied (Poisson effect), as well as the effect on the ionic conductivity that results from the presence of surfaces on this material; all under realistic dopant concentrations (10-20 \%) and high temperatures (1273-1673\,K). Also, this paper differs from previous ones \cite{KushimaAndYildizECST2009,KushimaAndYildiz_JMC2010,DeSouzaEtAl_EES2012,LiEtAl_PCCP2013} in that we only study small strain levels ($\epsilon$ $\le$ 0.021), that can be elastically accommodated in thin films. Finally, in our simulations, the Ce cation reduction reaction (Ce$^{4+}$ to Ce$^{3+}$), observed in some studies \cite{SuzukiEtAl_SSI2002}, is not allowed. This makes it possible to isolate the effects of strain on the \textit{ionic} conductivity only. \newline

%The slab method \cite{OliverEtAl_MSMSE1993,NolanEtAl_SS2005a} was employed because it allowed for the study of surface effects on thin films. Both bulk and slab cells were subjected to realistic strain levels ($\epsilon$ $\le$ 0.021). Such simulations isolate strain effects from others present at interfaces (e.g. dislocations). This work was carried out using MD calculations with accurate dipole-polarizable interatomic potentials derived directly from \textit{ab initio} calculations \cite{BurbanoEtAl_JPCM2011,BurbanoEtAl_CM2012}. Contrary to simulations which use static methods, these calculations were performed under realistic operating conditions, i.e.  high temperature and high dopant concentrations. 

\section{Methods}
\label{Methods}
\subsection{Interatomic Potential} 
The computational methods used in this study are well established and have been described elsewhere for doped ceria and other oxides \cite{MarrocchelliEtAl_JPCM2009,BurbanoEtAl_JPCM2011,BurbanoEtAl_CM2012,Marrocchelli_JPCM2010,WilsonEtAl_JPCM2004}, as well as fluoride systems \cite{HeatonEtAl_JPCB2006,SalanneEtAl_JFC2009}. The interatomic potential that was employed is known as the DIPole Polarizable Ion Model (DIPPIM) \cite{MaddenEtAl_JMST2006}. In this model the various ionic species are assigned their formal valence charges (Ce$^{4+}$, Y$^{3+}$, and O$^{2-}$) with the inclusion of the polarization effects that result from the induction of dipoles on the ions. The DIPPIM parameters were derived directly from hybrid density functional theory (h-DFT) calculations using the Heyd, Scuzeria, Ernzerhof (HSE06) functional \cite{HeydEtAl_JCP2003,KrukauEtAl_JCP2006}, as implemented in the VASP code \cite{PaierEtAl_JCP2006}. These static h-DFT calculations provided forces and dipoles of each ion, and were used to optimize the h-DIPPIM (HSE-DIPPIM) parameters. The procedure employed to generate a set of potentials for CeO$_2$-ZrO$_2$ solid solutions doped (or reduced) with trivalent RE cations (Sc$^{3+}$, Y$^{3+}$, Gd$^{3+}$, Sm$^{3+}$, Nd$^{3+}$, Ce$^{3+}$, La$^{3+}$) will be the subject of a future publication \cite{BurbanoEtAl_CeO22013}. The inclusion of a fraction of nonlocal Hartree-Fock exchange to standard DFT in functionals such as HSE is known to be necessary to correctly describe the electronic strucuture of lanthanide oxides such as ceria \cite{DaSilvaEtAl_PRB2007,GillenEtAl_PRB2013}. This is important in cases where highly correlated \textit{f}-electrons are present. However, for YDC this potential represents an improvement in terms of a closer agreement to experimental lattice constants \cite{DaSilvaEtAl_PRB2007}.  Table \ref{DIPPIMParams} presents the parameters for the h-DIPPIM interatomic potential used in this study. \newline

\begin{table*}[htb]
\begin{center}
\begin{tabular}{c c c c }
\noalign{\smallskip}\hline\noalign{\smallskip}
         	      & O$^{2-}$ -- O$^{2-}$ 		& Y$^{3+}$ -- O$^{2-}$ 	& Ce$^{4+}$ -- O$^{2-}$  \\
\hline
\\
$A^{ij}$ &  7.15    	& 118.0 	& 82.20 		 \\
$a^{ij}$ &  18.52     	& 1.38 	& 1.19 		 \\
$B^{ij}$ & 50000 	& 50000 	& 50000 		\\
$b^{ij}$ & 1.00   	& 1.50   	& 1.55   		\\
\\
\hline
\\
$C_6^{ij}$ & 83  	& 21  	& 47  		\\
$C_8^{ij}$ & 1240 	& 264 	& 595 		 \\
$b_{disp}^{ij}$ & 1.30  & 1.60  	& 1.50 		\\
$c_{disp}^{ij}$ & 1.70  & 2.08  	& 1.96 		\\
\\
\hline
\\
$\alpha_{\rm{O}^{2-}}$   & 13.97   	& 	&\\
$\alpha_{\rm{Y}^{3+}}$   & 2.31   	& 	& \\
$\alpha_{\rm{Ce}^{4+}}$  & 5.86   	& 	& \\
\\
\hline
\\
$b_D^{\rm{O}^{2-} \;-\; \rm{O}^{2-}}$ 		& 2.18     	& &  \\
$c_D^{\rm{O}^{2-} \;-\; \rm{O}^{2-}}$ 		& 3.03 	&  & \\
$b_D^{\rm{O}^{2-} \;-\; \rm{Y}^{3+}}$ 	&1.47 &  $b_D^{\rm{Y}^{3+} \;-\; \rm{O}^{2-}}$ 	& 1.47	\\
$c_D^{\rm{O}^{2-} \;-\; \rm{Y}^{3+}}$ 	&1.08 &  $c_D^{\rm{Y}^{3+} \;-\; \rm{O}^{2-}}$  	& -0.60	\\
$b_D^{\rm{O}^{2-} \;-\; \rm{Ce}^{4+}}$  	& 1.75	&  $b_D^{\rm{Ce}^{4+} \;-\; \rm{O}^{2-}}$ 	& 1.75    \\
$b_D^{\rm{O}^{2-} \;-\; \rm{Ce}^{4+}}$ 	& 1.85	&  $b_D^{\rm{Ce}^{4+} \;-\; \rm{O}^{2-}}$ 	& 0.17     \\
\\
\hline
\end{tabular}
\caption{Parameters for the h-DIPPIM potential. All values are in atomic units.}
\label{DIPPIMParams}
\end{center}
\end{table*}

\subsection{MD Simulation Details}

\begin{figure*}[htbp]
\begin{center}
\includegraphics[scale=0.1]{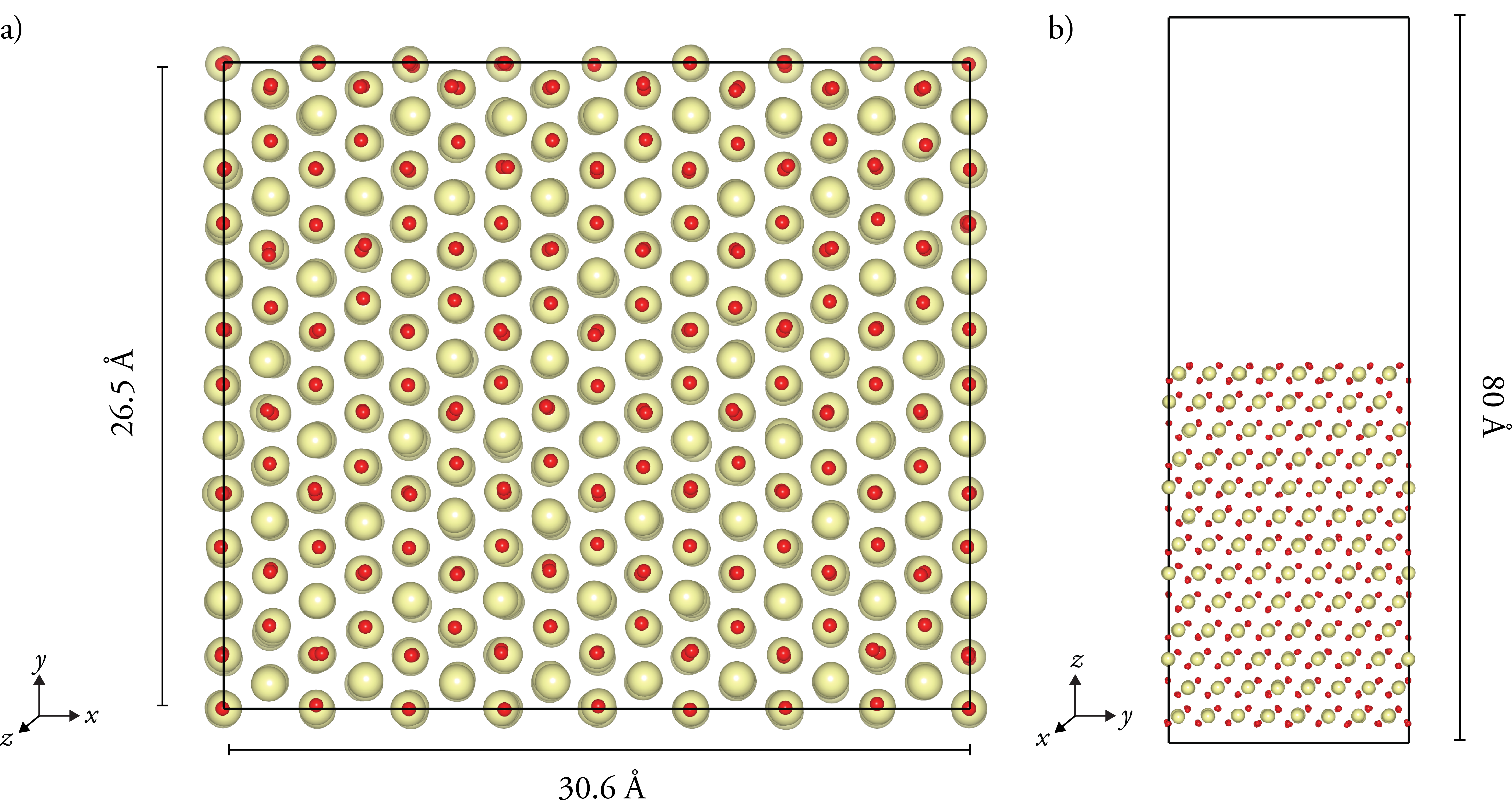}
\end{center}
\caption{a) CeO$_{2}$ slab \textit{xy}-plane b) Slab \textit{yz}-plane }
\label{Slab-Planes}
\end{figure*}

All Molecular Dynamics simulations were performed with an in-house code (PIMAIM) which uses three-dimensional periodic boundary conditions. The ionic conductivities of three different dopant concentrations were studied, namely Ce$_{1-x}$Y$_{x}$O$_{2-x/2}$ where $x = 0.08,0.12,0.18$. This choice of values originated from our previous work, where it was found that the conductivity maxium in YDC is at $x = 0.12$ \cite{BurbanoEtAl_CM2012}. The other two values for the dopant concentration were included in order to investigate if straining YDC causes a shift in the position of the conductivity maximum. As was the case in previous studies, the dopant cations and their corresponding charge compensating vacancies were distributed at random within their respective sublattices. Short MD simulations were carried out for each concentration at constant temperature and pressure (NPT ensemble) \cite{MartynaEtAl_JCP1994} in order to obtain the equilibrium lattice constants at three temperatures, T = 1673\,K, 1473\,K and 1273\,K. Isotropic tensile strain was simulated by expanding the axes of 6 x 6 x 6 fluorite supercells (henceforth referred to as bulk) under constant temperature and volume conditions (NVT ensemble). Each bulk simulation cell was subjected to three different strain levels $\epsilon_{x} = \epsilon_{y} = \epsilon_{z} = \{0.007, 0.014, 0.021\}$ for each temperature. YDC thin films of the same concentrations as bulk were simulated using slabs that exposed the (111) surface which has been found to be the most stable for ceria both in experiments and simulations \cite{NorenbergAndBriggs_SS1999,NolanEtAl_SS2005a}. The size of the simulation cells along \textit{x} and \textit{y} were $\sim$\,30\AA\ and $\sim$\,26\AA\ (Figure \ref{Slab-Planes}a), respectively, depending on the dopant concentration and temperature. The length along the \textit{z}-axis for the slab simulation cells was 80\,\AA. This corresponded to a vacuum gap of $\sim$ 40\,\AA\ depending on the dopant concentration and temperature (Picture \ref{Slab-Planes}b). The vacuum gap is necessary in order to avoid interactions between periodic images of the slabs along this direction. By using slabs with these dimensions it was possible to have similar numbers of ions of each species as in the bulk YDC cells. Table \ref{table:concs} reports the number of atoms from each species in these systems.  The simulation of anisotropic biaxial strain on YDC was performed by applying every combination of strain to the slabs, where $\epsilon_{x}, \epsilon_{y} = \{0.007, 0.014, 0.021\}$ (Picture \ref{Slab-Planes}a). These simulations were run in the NVT ensemble resulting in the lattice vectors along the \textit{x} and \textit{y} directions of the slabs being fixed while relaxation was allowed along the direction normal to the slab surface (\textit{z}-axis) in order to account for the Poisson effect (Picture \ref{Slab-Planes}b). 

\begin{table}[htb]
\begin{center}
\begin{tabular}{| c | c c c | c c c |}
\hline
  &   & Bulk  &   &  & Slab  &          \\
$x$ in Ce$_{1-x}$Y$_{x}$O$_{2-x/2}$ & Ce & Y & O & Ce & Y & O	 \\
\hline
0.08		& 796	& 68 & 1694 & 766 & 66 & 1631 \\
\hline	 
0.12 			& 762 & 102 & 1677 & 734 & 98 & 1615 \\
\hline	
0.18 			& 710 & 154 & 1651 & 684 & 148 & 1590 \\
\hline
\end{tabular}
\caption{Number of ions from each species in the bulk and slab simulation cells for each concentration of YDC.}
\label{table:concs}
\end{center}
\end{table}

The use of the DIPPIM interatomic potential made it possible to perform the long simulations required to measure the diffusion coefficients of these systems, which in the case of the lowest temperature considered was 600\,ps with a timestep of 1\,fs for all calculations. Coulombic and dispersion interactions were handled using Ewald summations, while the short-range part of the potential was truncated at 12.7\,\AA. 

\section{Results and discussion}
\label{Results}

\subsection{Bulk YDC conductivity}
\label{bulkcond}

%In a series of recent studies by the authors it was shown that the DIPPIM potential used in this work is able to account for the structural and conductivity properties of YDC \cite{BurbanoEtAl_JPCM2011,BurbanoEtAl_CM2012}. The newly parameterized DIPPIM potential used herein differs from our previous work in that h-DFT calculations (HSE) \cite{HeydEtAl_JCP2003,KrukauEtAl_JCP2006}, rather than Local Density Approximation (LDA)\cite{Hafner_JCC2008,AnisimovEtAl_JPCM1997,AnderssonEtAl_PhysRevB2007}, were used in the potential derivation process. The inclusion of a fraction of nonlocal Hartree-Fock exchange to standard DFT in functionals such as HSE is known to be necessary to correctly describe the electronic strucuture of lanthanide oxides such as ceria \cite{DaSilvaEtAl_PRB2007,GillenEtAl_PRB2013}. This change is only expected to have an inpact in cases where highly correlated \textit{f}-electrons are present. However, for YDC the new potential represents an improvement in terms of a closer agreement to experimental lattice constants \cite{DaSilvaEtAl_PRB2007}. The details of this potential's parameterization and the simulation of other ceria/zirconia systems with common oxygen terms will be the subject of future publications \cite{BurbanoEtAl_CeO22013}.  

\begin{figure}[htbp]
\begin{center}
\includegraphics[scale=0.40]{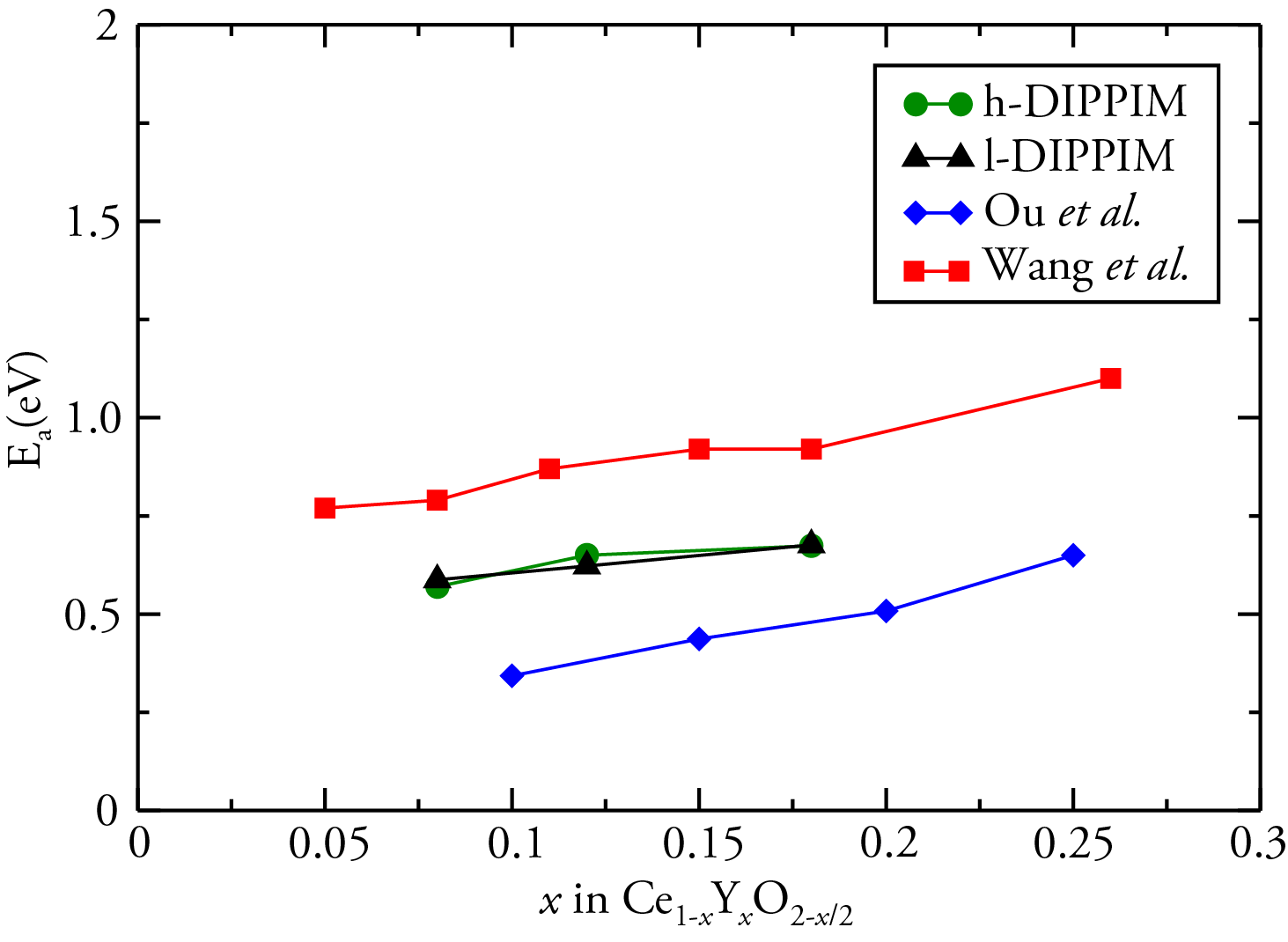}
\end{center}
\caption{Bulk activation energies, E$_{a}$ (eV), calculated using the h-DIPPIM (green circles) and l-DIPPIM potentials (black triangles. Experimental bulk activation energies from Ou \textit{et al.} \cite{OuEtAl_AM2006,Changrong_SOFC_Book} (blue diamonds) and Wang \textit{et al.} \cite{WangEtAl_SSI1981} (red squares)}
\label{fig:EaBulk}
\end{figure}

Figure \ref{fig:EaBulk} presents the activation energies, E$_{a}$ (eV), for vacancy migration in bulk YDC obtained from DIPPIM simulations using both the current parameters (Table \ref{DIPPIMParams}) and those from our previous work \cite{BurbanoEtAl_JPCM2011}, as well as, from experimental data available in the literature. The E$_{a}$ values predicted by the h-DIPPIM potential  were calculated from Arrhenius plots of the ionic conductivity in the temperature range 1273\,K -- 1673\,K (NVT ensemble) for each of the three concentrations considered in this study. The previous LDA-DIPPIM (l-DIPPIM) work used 4 $\times$ 4 $\times$ 4 cells under NPT conditions \cite{BurbanoEtAl_JPCM2011} and a wider temperature range (1073\,K -- 2073\,K). As expected, both sets of computational data are very similar despite being parameterized using different DFT functionals (h-DIPPIM shown as green circles and l-DIPPIM as black triangles). Moreover, the values from the simulations are in good agreement with the range of available experimental results (blue diamonds \cite{OuEtAl_AM2006} and red squares \cite{WangEtAl_SSI1981}). Table \ref{table:LDA-HSE} contains the DIPPIM calculated bulk $\sigma_{i}$ (S\,cm$^{-1}$) as a function of dopant concentration at 1473\,K. Both the h-DIPPIM and the l-DIPPIM potentials predict $x = 0.12$ to be the ionic conductivity maximum in YDC. The same level of agreement was observed between the potentials for all temperatues studied. This indicates that the h-DIPPIM potential maintains the previous potential's ability to account for the properties of YDC, but with the added advantages discussed above (Section \ref{Methods}). As a result, all subsequent parts of this study were performed using the h-DIPPIM potential. \newline

\begin{table}[htb]
\begin{center}
\begin{tabular}{| c | c c c c |}
\hline
     & & & $\sigma_{i}$ (S\,cm$^{-1}$)  &           \\
$x$ in Ce$_{1-x}$Y$_{x}$O$_{2-x/2}$ & & l-DIPPIM &  h-DIPPIM &\\
\hline
0.08		& & 0.110 & 0.125 & \\
\hline	 
0.12 	& & 0.132 & 0.136 & \\
\hline	
0.18 	& & 0.125 & 0.122 & \\
\hline			
\end{tabular}
\caption{Bulk ionic conductivities, $\sigma_{i}$ (S\,cm$^{-1}$), at 1473\,K from LDA (l-DIPPIM) \cite{BurbanoEtAl_JPCM2011,BurbanoEtAl_CM2012} and HSE (h-DIPPIM) DIPPIM potentials (present work).} 
\label{table:LDA-HSE}
\end{center}
\end{table}

\begin{figure}[htbp]
\begin{center}
\includegraphics[scale=0.40]{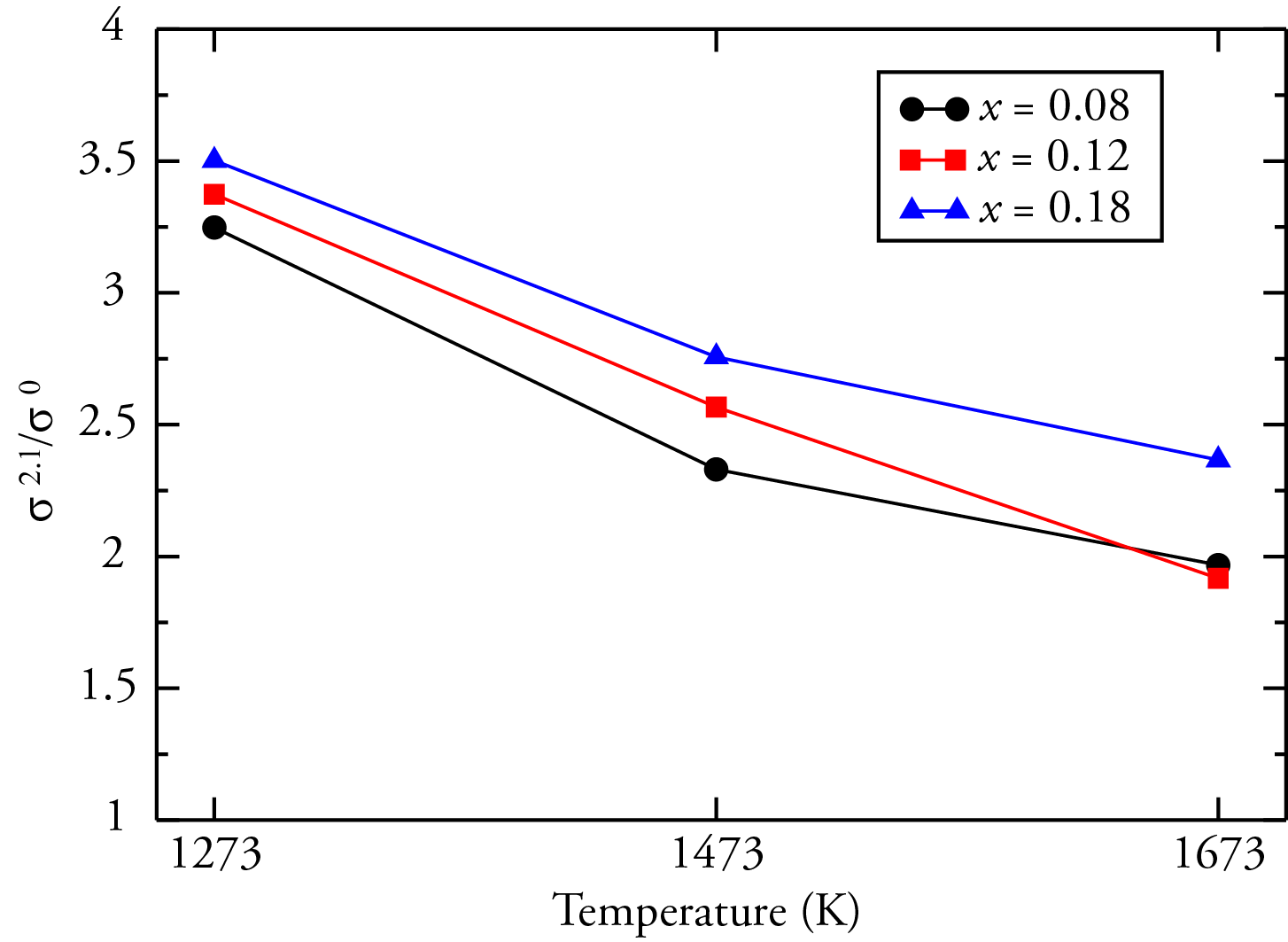}
\end{center}
\caption{Ionic conductivities of bulk Ce$_{1-x}$Y$_{x}$O$_{2-x/2}$ when $\epsilon = 0.021$ relative to $\epsilon = 0.000$ ($\sigma^{2.1}/\sigma^{0}$) as a function of temperature. The different concentrations are indicated as $x$ = 0.08 (black dots), 0.12 (red squares) and 0.18 (blue triangles).}
\label{BulkConds-strain}
\end{figure}

Application of isotropic strain to bulk YDC caused an increase in the material's ionic conductivity, $\sigma_{i}$. Figure \ref{BulkConds-strain} presents $\sigma_{i}$ of bulk Ce$_{1-x}$Y$_{x}$O$_{2-x/2}$ when $\epsilon = 0.021$ relative to $\epsilon = 0.000$ ($\sigma^{2.1}/\sigma^{0}$) as a function of temperature. The conductivity maximum remains at $x$ = 0.12 (red squares) for all the temperatures considered, however the highest increase in $\sigma_{i}$ with strain (3.50\,$\times$) is observed for $x$ = 0.18 (blue triangles) at 1273\,K. These results also show that the impact of strain on $\sigma_{i}$ increases as the temperature decreases from 1673\,K to 1273\,K. This effect can be explained in terms of the decrease in the activation energy of vacancy migration, E$_{a}$ (eV) (Table \ref{table:bulkEa}) due to the activated nature of the conduction mechanism. The magnitude of the changes observed in E$_{a}$ for YDC is similar to those reported by De Souza \textit{et al.} \cite{DeSouzaEtAl_EES2012}, who found a decrease of up to 40\% for CeO$_{2}$ at $\epsilon \sim 0.02$. Lower temperatures were not included in this study as they are more difficult to simulate because they require longer MD trajectories (in the order of several ns of simulation time). Unfortunately, isotropic strain is difficult to realize experimentally and is presented here due to its theoretical interest. As a consequence, the changes in conductivity experienced by YDC under these conditions will be used only as a benchmark when compared to those of anisotropically strained slabs in the next section. 

\begin{table}[htb]
\begin{center}
\begin{tabular}{| c | c c c c |}
\hline
  &   & & E$_{a}$ (eV) for $\epsilon$  &           \\
$x$ in Ce$_{1-x}$Y$_{x}$O$_{2-x/2}$ & 0.000 & 0.007 & 0.014 & 0.021 \\
\hline
0.08		& 0.570 &0.487 [-14.6] & 0.397 [-30.4] & 0.338 [-40.7] \\
\hline	 
0.12 		&0.650 & 0.580 [-10.8] & 0.554 [-14.8] & 0.392 [-39.6] \\
\hline	
0.18 		& 0.674 & 0.651 [-3.3] & 0.567 [-15.9] & 0.493 [-26.8]  \\
\hline	
\end{tabular}
\caption{Activation energies, E$_{a}$ (eV), as a function of strain, $\epsilon$, for bulk YDC.  The values in square brackets correspond to the percentage change with respect to the activation energy without strain for a given concentration.} 
\label{table:bulkEa}
\end{center}
\end{table}

\subsection{YDC slabs: surface effects}
\label{slabcond}

\begin{figure}[htbp]
\begin{center}
\includegraphics[scale=0.40]{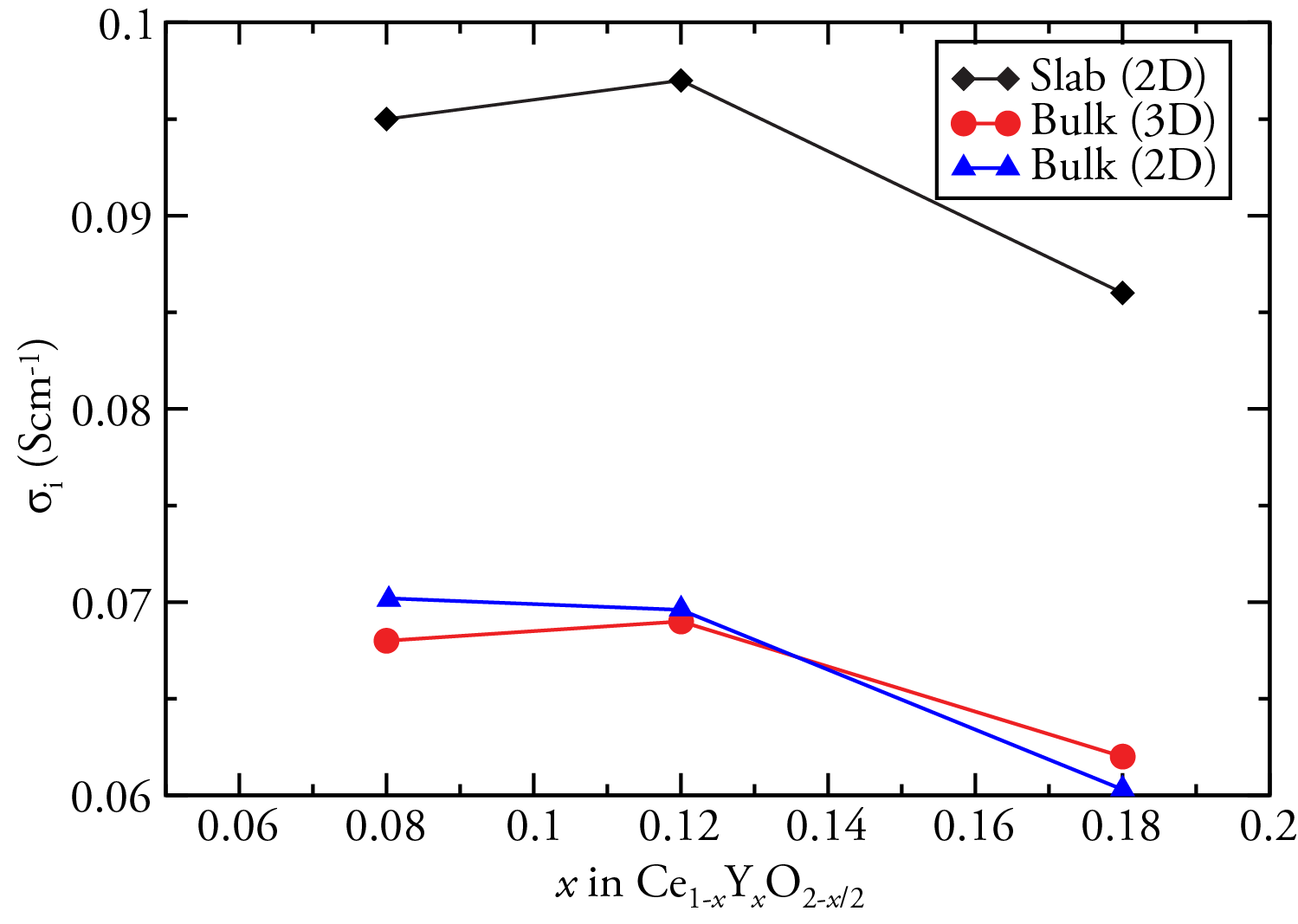}
\end{center}
\caption{2D (lateral) ionic conductivities, $\sigma_{i}$ (S\,cm$^{-1}$), for unstrained bulk (blue triangles) and slabs (black diamonds) as a function of dopant concentration at 1273\,K. Bulk 3D ionic conductivities are shown as red dots.}
\label{UnstrBulkVSlab-sig}
\end{figure}

The YDC slabs discussed in this section serve as models for the study of ionic conductivity of epitaxially grown thin films, in particular, coherently grown films where the lattice mismatch does not exceed $\epsilon$ = 0.03. Figure \ref{UnstrBulkVSlab-sig} presents the 2D ionic conductivities, $\sigma_{i}$ (S\,cm$^{-1}$), at 1273\,K for unstrained YDC as a function of dopant concentration for bulk (blue triangles) and slabs (black diamonds). These conductivities were calculated from the diffusion coefficients along the \textit{x} and \textit{y} directions only (D$_{xy}$ - 2D lateral diffusion). The MD simulations reveal that, for zero strain, the conductivities of the slabs are 50\% higher than those of bulk YDC of the same concentration. This pattern remains unaffected when all three components of the diffusion coefficient (D$_{x}$, D$_{y}$, D$_{z}$) are included in the calculation of the conductivity of bulk YDC and when this value is obtained from the average of the conductivities along the \textit{xy}, \textit{yz}, \textit{xz} planes (3D conductivity shown as red dots). To the authors' knowledge, this is the first time that such a behaviour is observed in YDC. Apart from the increase in the conductivity, the behaviour of the two systems is  similar, with the highest conductivity achieved for a dopant concentration of $x$ = 0.12.

\begin{figure}[htbp]
\begin{center}
\includegraphics[scale=0.45]{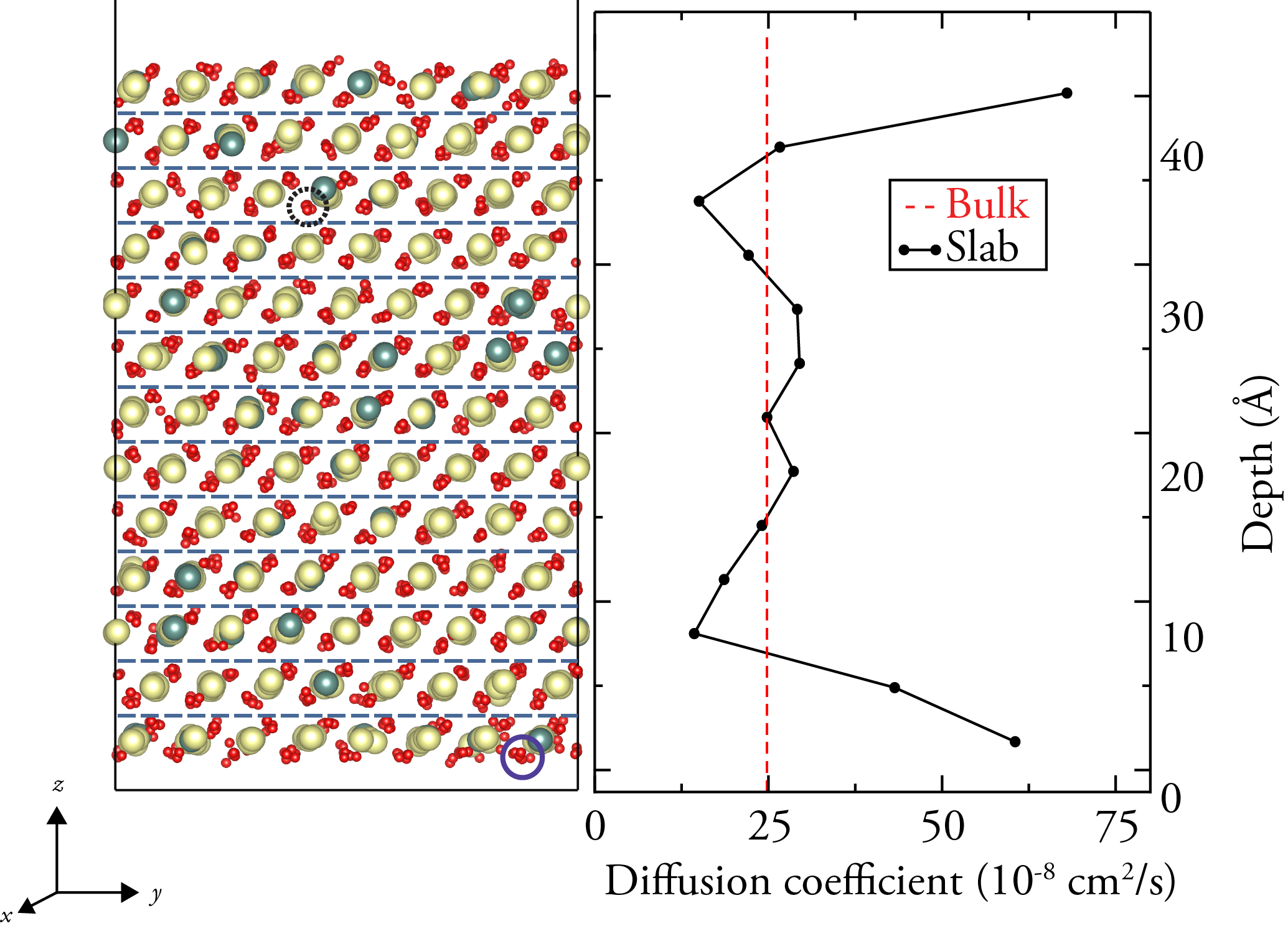}
\end{center}
\caption{3D diffusion coefficients for each atomic layer along the \textit{z}-axis for Ce$_{0.88}$Y$_{0.12}$O$_{1.94}$ at 1273\,K (black dots). The vertical red dashed line indicates the bulk conductivity at the same dopant concentration and temperature. Attention is drawn to the high degree of disorder in anion sublattice on surface layers (solid blue circle) compared to the ordered subsurface layers (dashed black circle).}
\label{DiffCoeff-Depth}
\end{figure}

The MD technique makes it possible to perform a detailed examination of the conductivity within each atomic layer along the \textit{z}-axis of the slabs (delimited by horizontal blue dashed lines on the left in Figure \ref{DiffCoeff-Depth}). The diffusion coefficients for each of these 13 layers of Ce$_{0.88}$Y$_{0.12}$O$_{1.94}$ at 1273\,K are presented on the right hand side of Figure \ref{DiffCoeff-Depth} as a function of depth (black dots aligned to their corresponding layers). It is immediately clear that the layers with exposed surfaces show a significantly higher conductivity ($\sim$\,4\,$\times$) than the bulk region of the slab, with the lowest conductivities seen in the subsurface layers.  The diffusion coefficient for bulk YDC of the same concentration as the slab is shown for comparison (vertical red dashed line). The increased conductivity in the surface regions is likely due to the more disordered atomic configurations observed in these layers. This higher degree of disorder is evident when the positions of the oxide anions on the surface regions (solid blue circle in Figure \ref{DiffCoeff-Depth}) are compared to those in the low conductivity subsurface (dashed black circle in Figure \ref{DiffCoeff-Depth}). \newline

%The fact that the diffusion coefficient in the middle of the slab (layer 7) behaves as bulk indicates the need for slightly thicker slabs in future simulations. \newline

\subsection{YDC slabs: strain effects}

\begin{figure*}[htbp]
\begin{center}
\includegraphics[scale=0.45]{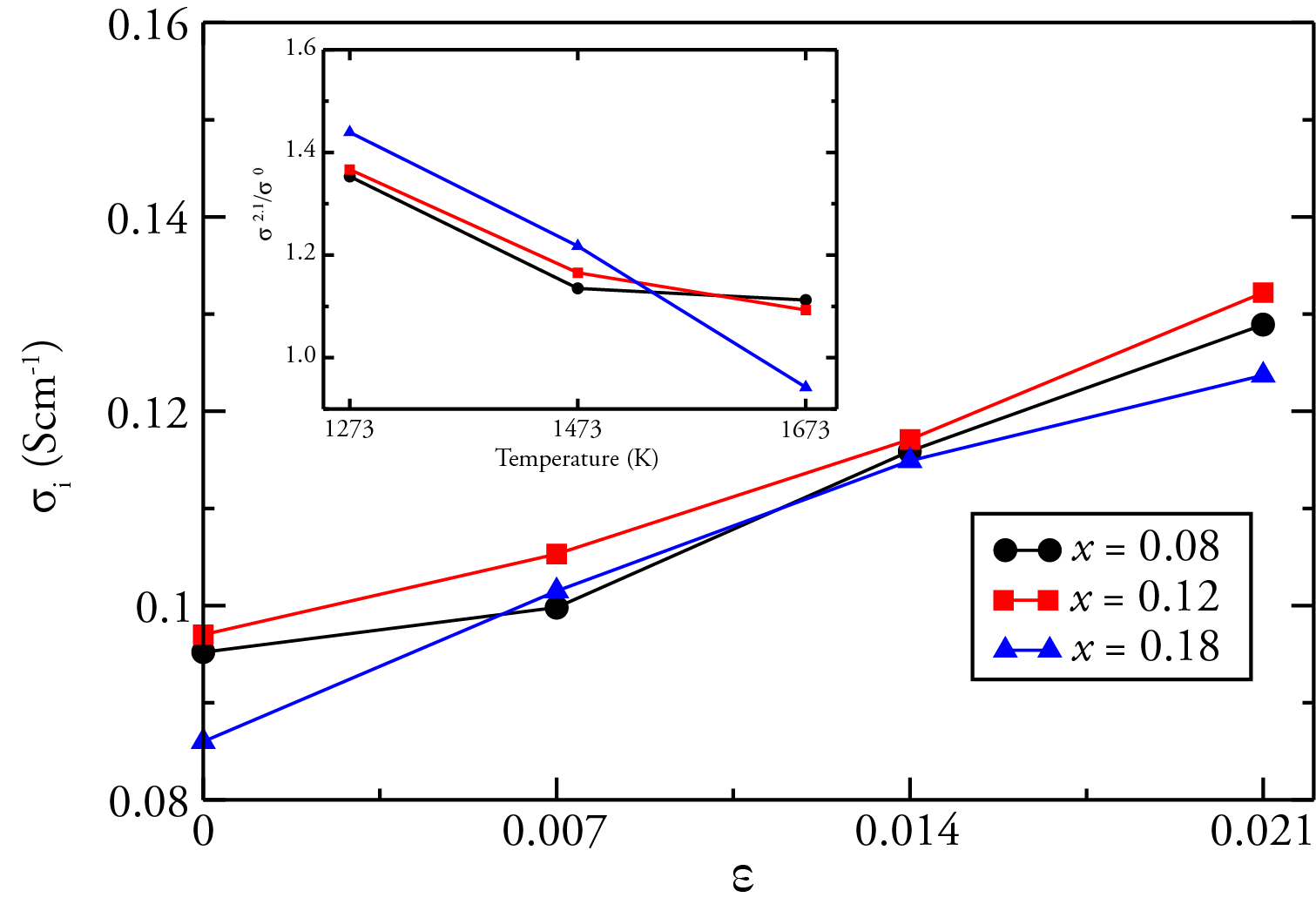}
\end{center}
\caption{Ionic conductivities of Ce$_{1-x}$Y$_{x}$O$_{2-x/2}$ slabs at 1273\,K as a function of strain. The different concentrations are indicated as $x$ = 0.08 (black dots), 0.12 (red squares) and 0.18 (blue triangles). Inset: 2D ionic conductivities of Ce$_{1-x}$Y$_{x}$O$_{2-x/2}$ slabs when $\epsilon_{x} = \epsilon_{y}$ = 0.021 relative to $\epsilon_{x} = \epsilon_{y}$ = 0.000 ($\sigma^{2.1}/\sigma^{0}$) as a function of temperature. }
\label{SlabConds-strain}
\end{figure*}

All combinations of strains, where $\epsilon_{x}, \epsilon_{y} = \{0.007, 0.014, 0.021\}$, were applied to the YDC slabs. However, it was found that, although the conductivity tended to increase whenever there was strain along either direction, the change was generally larger when $\epsilon_{x}$ = $\epsilon_{y}$ (which is to be expected since YDC adopts a cubic fluorite structure). For this reason, we will only report strain where  $\epsilon_{x}$ = $\epsilon_{y}$ for the slabs. In Figure \ref{SlabConds-strain} we report the conductivity vs strain for Ce$_{1-x}$Y$_{x}$O$_{2-x/2}$ slabs at 1273\,K, where $x$ = 0.08 (black dots), 0.12 (red squares) and 0.18 (blue triangles). The plots show that $\sigma_{i}$ increases with strain for all concentrations of YDC, and just as in the bulk system, $x$ = 0.12 displays the highest conductivity. The inset shows the change in conductivity of the same YDC systems when $\epsilon_{x} = \epsilon_{y}$ = 0.021 with respect to the unstrained slabs ($\sigma^{2.1}/\sigma^{0}$) for all temperatures under study. As was seen in bulk YDC (Figure \ref{BulkConds-strain}), applying strain to YDC causes a larger increase in conductivity at lower temperatures. In the case of slabs, this change corresponds to a 1.44\,$\times$ rise with respect to the unstrained slab at the same temperature for $x$ = 0.18 (\textit{cf.} the increase for isotropically strained bulk was 3.5\,$\times$). This increase may appear modest when compared to the orders of magnitude seen in some of the YSZ literature \cite{Garcia-BarriocanalEtAl_Science2008}, but it is in very good agreement with the latest reports on ceria and zirconia-based electrolytes \cite{SannaEtAl_AFM2009,SannaEtAl_Small2010,PerkinsEtAL_AFM2010,PergolesiEtAl_ACSN2012}. Our findings are in also reasonable agreement with previous calculations \cite{DeSouzaEtAl_EES2012,KushimaAndYildizECST2009,KushimaAndYildiz_JMC2010}, when the differences in strain values and temperatures are factored in. The activation energies, E$_{a}$ (eV), of the YDC slabs (Table \ref{table:slabEa}) decrease as the strain increases, nevertheless, these changes are not as pronounced as was the case of bulk YDC (Table \ref{table:bulkEa}), especially when $\epsilon$ = 0.021. Again, since the ionic conductivity is an activated process, the conductivity enhancement is expected to be higher at lower temperatures, as observed by Kushima and Yildiz for YSZ \cite{KushimaAndYildizECST2009,KushimaAndYildiz_JMC2010}. \newline

\begin{table*}[htb]
\begin{center}
\begin{tabular}{| c | c c c c |}
\hline
  &   & & E$_{a}$ (eV) for $\epsilon_{x}$	= $\epsilon_{y}$  &           \\
$x$ in Ce$_{1-x}$Y$_{x}$O$_{2-x/2}$ & 0.000 & 0.007 & 0.014 & 0.021 \\
\hline
0.08		& 0.511 &0.479 [-6.3] & 0.477 [-6.6] & 0.419 [-18.1] \\
\hline	 
0.12 		&0.608 & 0.528 [-13.2] & 0.510 [-16.1] & 0.504 [-17.1] \\
\hline	
0.18 		& 0.691 & 0.643 [-6.9] & 0.588 [-14.9] & 0.499 [-27.7]  \\
\hline	
		
\end{tabular}
\caption{Activation energies, E$_{a}$ (eV), as a function of strain, $\epsilon$, for YDC slabs. The values in square brackets correspond to the percentage change with respect to the activation energy without strain for a given concentration.} 
\label{table:slabEa}
\end{center}
\end{table*}

The ionic conductivities of the YDC slabs manifest a high degree of anisotropy, which is evidenced when the diffusion coefficients , D (10$^{-8}$\,cm$^{2}$/s), are decomposed into the individual components along each Cartesian direction. The values reported in Table \ref{Diff-aniso} correspond to the diffusion coefficients along the \textit{x} (D$_{x}$), \textit{y} (D$_{y}$) and \textit{z} (D$_{z}$) directions for biaxially strained Ce$_{0.88}$Y$_{0.12}$O$_{1.94}$ slabs and isotropically strained bulk (square brackets) of the same concentration. The results show that D$_{x}$ and D$_{y}$ increase more slowly in the slabs than they do in bulk when strain is applied. This anisotropy likely arises from the compression along the \textit{z} direction of the slabs (Poisson effect). The D$_{z}$ values of our simulation slabs are constrained by the non-periodicity along this direction and they will ultimately tend to zero as a result, provided the simulation time is long enough. However, in the timescales presented here the results indicate that the relaxation perpendicular to the surface plane does not lead to a diffusion enhancement. 

\begin{table}[htb]
\begin{center}
\begin{tabular}{|c | c c c |}
\hline
$\epsilon_{x}$	= $\epsilon_{y}$ & D$_{x}$ (10$^{-8}$ cm$^{2}$/s) & D$_{y}$ (10$^{-8}$ cm$^{2}$/s) & D$_{z}$ (10$^{-8}$ cm$^{2}$/s) \\
\hline	 
0.000 		& 33.5 [23.0] & 33.7 [26.8] & 23.3 [24.4] \\
\hline	
0.007 		& 38.1 [42.9] & 37.5 [42.4] & 24.5 [39.9] \\
\hline	
0.014 		& 40.5 [59.2] & 44.1 [62.5] & 25.4 [59.1] \\
\hline	
0.021 		& 49.3 [89.8] & 47.1 [89.9] & 26.7 [87.0] \\
\hline	
\end{tabular}
\caption{Slab diffusion coefficients along each Cartesian axis for Ce$_{0.88}$Y$_{0.12}$O$_{1.94}$ at 1273\,K under different strain, $\epsilon$, levels. Isotropically strained bulk diffusion coefficients are shown in square brackets for comparison.} 
\label{Diff-aniso}
\end{center}
\end{table}

\section{Conclusions}
\label{Conclusions}
This work attempts to rationalize the impact of strain in epitaxially strained RE-doped ceria. The simulations used accurate dipole-polarizable interatomic potentials derived directly from \textit{ab initio} calculations. The simulation conditions were realistic from the point of view of the strains applied, dopant cation concentrations and the temperatures employed. Also, for the first time, the slab method was applied to this problem which allowed for a more realistic representation of biaxially strained systems. In addition, the results from the slab calculations illustrated the effects of surfaces on the ionic conductivity. The results obtained from our calculations indicate that there is a clear enhancement of the conductivity in the surface region of a thin film. This resulted in our films being $\sim$ 50\% more conducting than the corresponding bulk sample, for zero strain. This effect is related to the more disordered atomic configurations observed in the surface layers and the presence of under-coordinated atoms. When strain is applied, either isotropically (bulk) or biaxially (slab), a moderate enhancement in the ionic conductivity is observed, for the studied temperature and strain ranges. Biaxially strained thin films are found to be less conductive than isotropically strained YDC, which has been ascribed to the highly anisotropic diffusion mechanism observed in the thin films. Our findings confirm those from recent experimental studies which employed films of high crystallographic quality and found only a limited impact on the ionic conductivity from tensile strain \cite{SannaEtAl_Small2010,PergolesiEtAl_ACSN2012}. 
\newline

\begin{acknowledgements}
MB acknowledges the HPC-EUROPA2 project (project number: 228398) with the support of the European Commission Capacities Area - Research Infrastructures Initiative. Calculations were performed on the JADE supercomputer maintained by CINES, Stokes (allocations tcche026b and tcche031b) maintained by ICHEC and Kelvin maintained by TCHPC. GWW acknowledges Science Foundation Ireland  (grant numbers 09/RFP/MTR2274 and 08/RFP/MTR1044). DM wishes to thank the Government of Ireland for an EMPOWER Postdoctoral Fellowship and is thankful to Profs. Harry Tuller and Bilge Yildiz for his recent appointment as visiting scientist at MIT.
\end{acknowledgements}

\newpage

\bibliographystyle{spphys}       % APS-like style for physics
\bibliography{BIB}   % name your BibTeX data base

\end{document}